\documentclass[useAMS,usenatbib]{mn2e}                                                             
\usepackage[totalwidth=515pt,totalheight=660pt,left=1.4cm,right=1.4cm]{geometry}              
\usepackage{graphicx,amssymb,color}                                                         
\usepackage[normalem]{ulem}                                                                    
\input psfig 

\title[Forming WD discs II: shrinking]
{Formation of planetary debris discs around white dwarfs II:
Shrinking extremely eccentric collisionless rings}
\author[Veras, Leinhardt, Eggl \& G\"{a}nsicke]{
Dimitri Veras$^{1}$\thanks{E-mail:d.veras@warwick.ac.uk},
Zo\"{e} M. Leinhardt$^{2}$,
Siegfried Eggl$^{3}$,
Boris T. G\"{a}nsicke$^{1}$
\\
$^{1}$Department of Physics, University of Warwick, Coventry CV4 7AL, UK
\\
$^{2}$The School of Physics, University of Bristol, Bristol BS8 1TL, UK
\\
$^{3}$IMCCE Observatroire de Paris, UPMC, 77 Av. Denfert-Rochereau, 75014 Paris, France
}

\begin{document}

\date{Accepted 2015 May 22. Received 2015 May 20; in original form 2015 February 25}

\pagerange{\pageref{firstpage}--\pageref{lastpage}} \pubyear{2015} 

\maketitle

\label{firstpage}

\begin{abstract}
The formation channel of the tens of compact debris discs which orbit white dwarfs (WDs) at a distance of one Solar radius remains unknown.  Asteroids that survive the giant branch stellar phases beyond a few au are assumed to be dynamically thrust towards the WD and tidally disrupted within its Roche radius, generating extremely eccentric ($e>0.98$) rings.  Here, we establish that WD radiation compresses and circularizes the orbits of super-micron to cm-sized ring constituents to entirely within the WD's Roche radius.  We derive a closed algebraic formula which well-approximates the shrinking time as a function of WD cooling age, the physical properties of the star and the physical and orbital properties of the ring particles.  The shrinking timescale increases with both particle size and cooling age, yielding age-dependent WD debris disc size distributions.
\end{abstract}


\section{Introduction}

Mounting discoveries of debris orbiting white dwarfs (WDs) presage increased scrutiny of post-main-sequence 
planetary systems.  The over 30 dusty discs \citep{zucbec1987,becetal2005,kiletal2005,reaetal2005,
faretal2009,baretal2014,beretal2014,rocetal2014} 
and 7 gaseous discs \citep{ganetal2006,ganetal2007,ganetal2008,gansicke2011,faretal2012,meletal2012,wiletal2014,manetal2015} so far 
detected all have radial extents of just about 1 Solar radius $\approx 0.005$ au $< 10^6$ km.  Such compact 
configurations are absent from main sequence planetary studies because they cannot exist; the discs would
be inside of the star!  Consequently, achieving an understanding of 
how these post-main-sequence discs are formed requires a different approach.

Protoplanetary discs orbiting young main-sequence stars form out of a collapsing 
stellar birth cloud, but WD discs must instead be formed from the tidal disruption of objects which encounter the WD 
\citep{graetal1990,jura2003,debetal2012,beasok2013}.  However, because of mass loss and stellar tides on the giant 
branch phases of stellar evolution, there exists a ``planet desert''.  Mass loss serves to push individual 
planets outward (sometimes to 
the interstellar medium; \citealt*{veretal2011,vertou2012,adaetal2013,veretal2014a}), and may cause multiple 
planets to become unstable \citep{dunlis1998,debsig2002,veretal2013a,voyetal2013,musetal2014,vergae2015}, leading to 
collisions, engulfment within the star, or escape from the system.  Concurrently, tidal effects between
the giant star and a close-in planet could overcome the outward evolution due to mass loss and cause the
star to swallow the planet \citep{viletal2009,kunetal2011,musvil2012,spimad2012,adablo2013,viletal2014}.  Therefore, 
we expect a void of planetary bodies within a few au of WDs.

Asteroids, comets, moons and planets are instead likely to exist beyond a few au in WD systems.  
Remnant exo-asteroid or exo-Kuiper 
belts can then dynamically interact with these planets, flinging asteroids towards the WD and eventual 
destruction \citep{bonetal2011,debetal2012,frehan2014}\footnote{Other potential reservoirs of disrupted 
material, such as exo-Oort clouds, are both compositionally \citep{zucetal2007} and dynamically  
\citep{veretal2014b,stoetal2015} unlikely to represent the primary source of the discs.}.  
The disruption of a single asteroid, which must 
be on a highly eccentric orbit in order to enter the disruption sphere (or Roche radius) of the WD, 
results in a highly eccentric ring 
\citep[][hereafter Paper I]{veretal2014c}.  By modelling the asteroid as a rubble pile composed
of equal-mass, equal-radii indestructible spheres, Paper I found that the resulting eccentric ring 
is collisionless.  The tidal disruption of multiple asteroids in quick succession might
instead produce collisional rings or discs.  However, here we treat the single asteroid (collisionless)
case as the next logical step to follow Paper I. The resulting eccentric ring maintains the same orbit as the
progenitor, but precesses due to general relativity.  Paper I found that the eccentricity $e$ of these 
orbits always satisfies $e > 0.98$,
with typical values of $e \sim 0.995$; an asteroid with a semimajor axis $a$ of 10 au
that skims a typical WD would satisfy $e \approx 0.99999$.  If the rings remain collisionless,
then without the influence of additional forces, they will never change shape.

However, observations of gaseous WD discs reveal circular or near-circular geometries.  Double-peaked
metal line emission profiles suggest $e \approx 0.02$ and $e \sim 0.2$ in the two most robustly-constrained 
systems \citep{ganetal2006,ganetal2008}.  The process of converting a highly eccentric orbit into a
nearly circular one, while drastically reducing the semimajor axis, has rarely been discussed in the context of WDs, and is 
the issue we address in this paper.  

To do so, we introduce radiative forces into the two-body problem, as each ring constituent can be considered
to be orbiting the WD unperturbed by any other body if the constituents are indeed collisionless.
We focus on radiation and do not consider effects from tidal dissipation in this work.
Regardless, radiation forces may dominate over tidal dissipation effects because the latter 
are likely to cause orbital changes of a ring particle on a much longer timescale.  The effects
of tidal dissipation in small and light ring constituents should be negligible, as can be deduced 
directly from the equations of motion (e.g. equation 8 of \citealt*{beanes2012}) or from relations
for the orbit-averaged semimajor axis and eccentricity damping timescales (e.g. equations 
10-11 of \citealt*{nagida2011}).  Both of those studies account for dynamical tides, a physical description
in which highly eccentric orbits may be treated.

Radiation, in the restricted, specific case of Poynting-Robertson drag, has become a crucial 
component of explaining accretion from compact circumstellar discs onto WDs
\citep{bocraf2011,rafikov2011a,rafikov2011b,metetal2012}.  \cite{wyaetal2014}
has constrained the possible size distributions of the accreting material
by fitting model parameters to observationally-inferred accretion rate distributions.
We know that accretion onto the WD occurs because associated 
with every dusty and gaseous disc are unmistakable signatures of photospheric metal pollution.
These metals cannot arise from stellar dredge-up nor from the
interstellar medium (ISM) due to the high accretion rate, the distribution of interstellar clouds
and the existence of metal-rich WDs that lack hydrogen 
\citep{aanetal1993,frietal2004,jura2006,kilred2007,ganetal2008,faretal2010}.  Here, we consider WD radiation 
in a different, earlier context, before a circular compact disc
has formed.  We are particularly motivated by the suggestion that Poynting-Robertson drag
plays an important role in the dynamics of highly-eccentric orbits \citep{voketal2001}, as
the tidally-disrupted debris from Paper I settled into such orbits.

In Section 2, we present our framework of perturbations due to WD radiation.  Sections 3 and 4
describe how radiation changes orbits on short and long timescales, respectively.  Based on these 
results, we derive the key formulae for orbital shrinkage in Section 5 before discussing the implications for the resulting ring structure due to collisions and the Yarkovsky effect in Section 6
and concluding in Section 7.

\section{General formulation} \label{PRDRAG}

We let $\vec{r}(t)$ represent the distance from the centre of the ring particle
to the centre of the WD, $\vec{v}(t)$ the velocity of the particle with 
respect to the centre of the WD, $m$ the particle's mass, $A$ the particle's
momentum-carrying cross-sectional area, $c$ the speed of light 
and $L(t)$ the time-varying luminosity of the WD.
Then the extra acceleration of the particle due to 
stellar radiation is \citep{veretal2015a}

\begin{equation}
\left(\frac{d\vec{v}}{dt}\right)_{\rm rad}
=
\frac{AL}{4\pi m c r^2}
\bigg[  
Q_{\rm abs}\mathbb{I} + Q_{\rm ref}\mathbb{I} + kQ_{\rm yar}\mathbb{Y}
\bigg]
\vec{\iota}
,
\label{finaldvdtp}
\end{equation}

\noindent{}where $Q_{\rm abs}$ is the absorption efficiency, $Q_{\rm ref}$
is the reflection efficiency, $Q_{\rm yar} = Q_{\rm abs} - Q_{\rm ref}$, 
$\mathbb{I}$ is the 3x3 unit matrix,
and $\mathbb{Y}$ is the 3x3 Yarkovsky matrix (given in equation
28 of \citealt*{veretal2015a}).  The thermal redistribution parameter
$k$ satisfies $0 \le k \le 1/4$.  The three terms in equation 
(\ref{finaldvdtp}) refer to the contributions from radiation which 
is absorbed, directly reflected and thermally emitted. 
The vector

\begin{equation}
\vec{\iota}
\equiv
\left(1 - \frac{\vec{v} \cdot \vec{r}}{cr} \right) \frac{\vec{r}}{r}
- \frac{\vec{v}}{c}
\label{PRBurns2}
\end{equation}

\noindent{}represents the relativistically-corrected direction of
incoming radiation.  

The above equations hold when the particle is at least a few
orders of magnitude larger than the wavelength of incoming radiation,
$\bar{\lambda}$. For WDs, $\bar{\lambda}$ changes with cooling age 
(time since the star became a WD).
Young ($\sim$4 Myr old), very hot WDs with temperatures of about 40000 K
will have a value of $\bar{\lambda}$ that is just 
under $1 \times 10^{-7}$ m.  Alternatively, old ($\sim$2 Gyr old) and
cold (6000 K) WDs have $\bar{\lambda} \approx 5 \times 10^{-7}$ m.
Effectively, throughout a WD's lifetime, $\bar{\lambda}$ varies by
less than an order of magnitude, and is always safely smaller than
a micron.

The contribution from the Yarkovsky term is negligible when
the particle size is larger than about 1 cm - 10 m \citep{veretal2015a}
and when the particle spins, such that some of the absorbed 
radiation is redistributed
before being emitted.  When the Yarkovsky effect is active,
it can represent the dominant contribution to the motion: the
resulting perturbation may eject an asteroid orbiting a $10^3 L_{\odot}$ evolved
star (such as an asymptotic giant branch star or very young WD) in 
just tens of million years. Here, we assume that the tidal
disruption of an asteroid into sub-metre-sized pieces occurs
on a much shorter timescale.  

Paper I suggests the number of orbits
required for the asteroids that were modeled to dissociate completely
into its individual constituents was tens to hundreds.  The 
pericentres of these asteroid orbits were within about 10\% of 
the WD's Roche radius, and the modeled asteroids were
strengthless rubble piles.  If the asteroids 
were instead internally more cohesive, then the disruption limits
would change.  Asteroids closer to the outer edge of the Roche 
radius may indeed disrupt
into sufficiently small bits on longer timescales, such that the 
Yarkovsky effect does play a significant role.

We discuss this possibility further in Section 6.  However, for the body 
of this paper, we continue the story begun in Paper I and consider
the subsequent evolution of the rings formed from disruption due to
close pericentre passages.  We assume that the constituents of 
these rings are small enough to be unaffected by the Yarkovsky effect, 
and we operate on this assumption for the remainder of Sections 2-5.  
Consequently, equation (\ref{finaldvdtp}) simplifies to

\begin{equation}
\left(\frac{d\vec{v}}{dt}\right)_{\rm rad}^{\mathbb{Y}=0}
=
\frac
{AL \mathcal{Q}}
{4\pi m c r^2}
\vec{\iota}
.
\label{finaldvdtp2}
\end{equation}

\noindent{}where

\begin{equation}
\mathcal{Q} \equiv Q_{\rm abs} + Q_{\rm ref}
.
\end{equation}

The luminosity of the WD is a steep function of the star's cooling 
age within the first hundreds of Myr.  We adopt the same dependence 
on cooling age as in equation (6) of \cite{bonwya2010} and re-express
their equation as

\begin{equation}
L(t) = 
3.26 L_{\odot} 
\left(\frac{M_{\star}}{0.6M_{\odot}}\right)
\left(0.1 + \frac{t}{{\rm Myr}} \right)^{-1.18}
,
\label{lum}
\end{equation}

\noindent{}which is applicable for about the first 9 Gyr of WD cooling (sufficient for our purposes).
Here $M_{\star}$ refers to the stellar mass.

We now determine how the radiative force from equation (\ref{finaldvdtp2}) reshapes
the debris rings.  First, we consider the perturbations on orbital timescales, and then model the 
cumulative (i.e. long-term, averaged, or secular) consequences.

\section{Unaveraged equations of motion}

The orbit of the particle and star is fixed unless acted upon by perturbations, such
as that from equation (\ref{finaldvdtp2}).  Based on equations from \cite{efroimsky2005}
and \cite{gurfil2007}, \cite{vereva2013a} outlined an algorithm
which can produce expressions in orbital elements for the time evolution 
of $a$ (semimajor axis), $e$ (eccentricity), $i$ (inclination), 
$\Omega$ (longitude of ascending node), $\omega,$ (argument of pericentre), 
$f$ (true anomaly), and the pericentre $q = a(1-e)$ and 
apocentre $\tilde{q} = a(1+e)$, in the perturbed two-body problem.  This technique 
eliminates all Cartesian components,
and does not make any assumptions about the orbital elements except that the motion
remains bounded.  This facet is important because the eccentricities we will be 
treating are near unity, and hence traditional disturbing functions expanded about
small eccentricities would not be applicable\footnote{\cite{veras2007} showed how
increasing the order of the expansion helps only until the Sundman criterion 
\citep{ferrazmello1994} is violated.}.

The complete equations of motion due to the perturbation from equation (\ref{finaldvdtp2})
have already been derived in \cite{veretal2015a}, and we do not repeat them here.
One quantity of particular interest not emphasized in that paper is the closeness of the 
orbital pericentre to the WD for an extremely eccentric orbit.  We obtain the time 
evolution of $q$ here by manipulating the equations for the
evolution of $a$ and $e$ in \cite{veretal2015a}, and find

\[
\left(\frac{dq}{dt}\right)^{\mathbb{Y}=0} 
=
-\frac{AL\mathcal{Q}\left(1+e\cos{f}\right)^2}{2\pi m\left(1+e\right)^{2}}
\times
\bigg[
\frac
{\sin{f} }
{2 n a^2 \sqrt{1-e^2}}
\left(\frac{1}{c}\right)
\]

\begin{equation}
\ \ \ \ \
+
\frac
{\sin^2\left({\frac{f}{2}}\right) \left[2 - e - e \cos{f}\right]}
{a \left(1-e^2\right)}
\left(\frac{1}{c^2}\right)
\bigg]
.
\label{qunpert}
\end{equation}

We need to know how much radiation will drag a particle located
at the pericentre of the orbit ($f=0$) towards (or away from) the WD relative to the particle's
unperturbed osculating location.  In the next section
we can obtain an order-of-magnitude estimate of the maximum 
value of this displacement by averaging equation (\ref{qunpert}) over 
an entire orbit.

\section{Averaged equations of motion} \label{avgPR}

In order to consider the cumulative effect of WD radiation
over many orbits, we analyze the averaged equations of motion.
We perform the averaging over the true 
anomaly such that for an arbitrary variable $\beta$,

\begin{equation}
\left\langle 
\frac{d\beta}{dt} 
\right\rangle
\equiv \frac{1}{2\pi} 
\int_{0}^{2\pi} \frac{d\beta}{dt} 
\frac{\left(1 - e^2\right)^{3/2}}{\left(1 + e \cos{f}\right)^2}
df
.
\end{equation}

\noindent{}We obtain \citep{veretal2015a}

\begin{equation}
\left\langle 
\left( \frac{da}{dt} \right)^{\mathbb{Y}=0}
\right\rangle
= 
-
\frac{AL\mathcal{Q} \left(2 + 3 e^2\right)}
{4 \pi m a c^2 \left(1 - e^2\right)^{3/2} }
,
\label{dadtavg}
\end{equation}

\begin{equation}
\left\langle 
\left( \frac{de}{dt} \right)^{\mathbb{Y}=0}
\right\rangle
= 
-
\frac{5 AL\mathcal{Q} e}
{8 \pi m a^2 c^2 \sqrt{1 - e^2} }
.
\label{dedtavg}
\end{equation}

\noindent{}Equations (\ref{dadtavg}) and (\ref{dedtavg}) agree with \cite{wyawhi1950},
and demonstrate that radiation decreases both the semimajor axis and eccentricity
over time.  Note that the $(1/c)$ terms vanish upon averaging. 

There is no long-term change in the argument or longitude of pericentre 

\begin{equation}
\left\langle 
\left(
\frac{d\omega}{dt}
\right)^{\mathbb{Y}=0}
\right\rangle
= 
\left\langle 
\left(
\frac{d\varpi}{dt}
\right)^{\mathbb{Y}=0}
\right\rangle
=
0
\label{domegadtavg}
\end{equation}

\noindent{}which importantly illustrates that the long-term precession of the orbit
is dictated by general relativity (in the absence of other, additional forces).  
Finally, we find that

\begin{equation}
\left\langle 
\left(
\frac{dq}{dt} 
\right)^{\mathbb{Y}=0}
\right\rangle
= 
-
\frac
{AL\mathcal{Q} \sqrt{1 - e^2} \left(4 - e\right)}
{8 \pi m a c^2 \left(1 + e\right)^2}
,
\label{qeq}
\end{equation}

\begin{equation}
\left\langle 
\left(
\frac{d\widetilde{q}}{dt} 
\right)^{\mathbb{Y}=0}
\right\rangle
= 
-
\frac
{AL\mathcal{Q} \sqrt{1 - e^2} \left(4 + e\right)}
{8 \pi m a c^2 \left(1 - e\right)^2}
,
\end{equation}

\noindent{}proving that Poynting-Robertson drag decreases both the pericentre and apocentre over time.

A closer look at equation (\ref{qeq}) can reveal by how much radiation causes the pericentre
to shrink per orbit.  Let $\Delta$ represent the actual close encounter distance minus
the distance predicted by Newtonian gravity with no additional forces per orbit.  We find

\[
\Delta \approx 
-0.364 \ {\rm m} \left(\mathcal{Q} 
\frac{\sqrt{1-e^2}\left(4-e\right)}{\left(1+e\right)^2} \right)
\left(\frac{a}{10 \ {\rm au} }\right)^{1/2}
\left(\frac{m}{10^{12} \ {\rm kg} } \right)^{-\frac{1}{3}} 
\]

\begin{equation}
\ \ \ \ \times
\left(\frac{M_{\star}}{0.6M_{\odot}} \right)^{-\frac{1}{2}} 
\left(\frac{\rho}{2 \ {\rm g/cm}^3 } \right)^{-\frac{2}{3}} 
\left(0.1 + \frac{t}{\rm Myr}  \right)^{-1.18}
\label{DelPR}
\end{equation}

\noindent{}where $\rho$ represents the particle density.
Hence, along a single orbit, radiation from a newly-born WD drags a body
towards the star by about 4 m at the pericentre.  For cooling ages of 1 Myr and 1 Gyr,
this value drops to about 0.4 m and 0.4 mm.  For highly eccentric orbits, these estimates
would be reduced by at least one order of magnitude. To place these values in context, consider 
that a typical WD radius is $\simeq 0.015 R_{\odot} = 10^4$ km 
\citep{hamsal1961,holetal2012,paretal2012} and the maximum possible extent of the 
WD disruption sphere is $\simeq 3.6 R_{\odot} = 2.5 \times 10^6$ km (Paper I).
Hence, over the lifetime of the Universe, this gradual pericentre drift is negligible
compared to the WD radius or its disruption sphere.

Further, we know that general relativity produces no such accumulation of inward
or outward drag despite causing a potentially relatively large deviation
from the Newtonian orbit at each pericentre passage.  The maximum
value of this deviation is about 9 km $\times (M_{\star}/M_{\odot})$, and the body is pushed towards the
star only when $e \lesssim 0.359$ \citep{veras2014}. 

By using both the averaged equations of motion (equations \ref{dadtavg}, \ref{dedtavg} and 
\ref{domegadtavg}) and a prescription
for the luminosity evolution (equation \ref{lum}), we can numerically determine the 
long-term orbital evolution of particles from WD radiation.  This approach requires
solving coupled differential equations, as we cannot find an explicit solution of the
general equations.  However, in the specific case of high eccentricity, we have
found an explicit solution, as described in the next section.

First, we numerically integrate these full averaged equations of motion (equations \ref{lum}, \ref{dadtavg},
\ref{dedtavg} and \ref{domegadtavg}) to determine the evolution for WD planetary systems.  Results 
of this integration are plotted as solid lines in Figure \ref{PRplots1} for bodies with radii ($R$) spanning
four orders of magnitude.  {\it Only for this size range ($10^{-5}-10^{-1}$ m) can we be sure that other radiative effects
do not come into play}.  Each panel illustrates a different WD cooling age $t_{\rm ini}$; older
WDs take longer to shrink orbits.  For simplicity, we adopted a characteristic WD mass ($M_{\star}$) of $0.6M_{\odot}$, 
characteristic body density ($\rho$) of $2$ g/cm$^3$ and $\mathcal{Q}=1$, for the integrations. 
Assuming $\mathcal{Q}=1$ means that no directed scattering or reemission occurs 
(i.e. $Q_{\rm ref} = Q_{\rm yar} = 0$), such that all radiation is absorbed and emitted 
omnidirectionally.  This assumption accurately reproduces the properties of small carbonaceous
dust particles with albedos of about 0.005.  Alternatively, typical asteroids with albedos between
0.1 and 0.3 would satisfy $1.1 < \mathcal{Q} < 1.3$.\footnote{The albedo distribution for main
belt asteroids is bimodal, with peaks at about 0.05 and 0.25.}

We also adopted one of 
the pericentre values used in Paper I to better demonstrate the link with that paper and to provide
analytically-motivated initial parameters.  In that respect, $q_{\rm min} = 0.0126R_{\odot} = 
5.86 \times 10^{-5}$ au is the radius of a typical
$0.6 M_{\odot}$ WD, and $q_{\rm max} = 2.73R_{\odot} = 0.013$ au is the maximum value of the disruption
sphere for that WD mass.

The figure demonstrates that for a wide range of WD cooling ages, all particles in this size range will 
maintain their original semimajor axes for the vast majority of their evolution, before their orbits suddenly shrink
to within the WD Roche radius.  Also plotted is the apocentre of the orbit.  For the majority of the 
particle's evolution, the particle maintains its initial eccentricity; circularization occurs only as 
the semimajor axis is shrinking drastically.  This behaviour is mirrored for particle evolution at
later WD cooling ages, although the collision timescale is increased, as will be shown in the next section.

\section{High eccentricity limit} \label{highePR}

Numerical simulations from the last subsection demonstrate that a particle on an initially highly eccentric 
orbit will remain so for the vast majority of its lifetime.  This result suggests that
considering the high-eccentricity limit of the equations might represent a good approximation
to the actual motion.  Our objective in this subsection is to express this motion 
by an explicit algebraic expression, which will facilitate future study.

We perform series expansions about $\epsilon = 1 - e$, where $\epsilon \ll 1$, and denote
the result with a tilde.  We obtain

\[
\widetilde{
\left\langle 
\left(
\frac{da}{dt} 
\right)^{\mathbb{Y}=0}
\right\rangle
}
= 
-
\frac
{5AL\mathcal{Q}}
{8\sqrt{2} \pi m a c^2 \left(1 - e\right)^{3/2}}
\]

\begin{equation}
\ \ \ \ \
+
\frac
{9AL\mathcal{Q}}
{32\sqrt{2} \pi m a c^2 \left(1 - e\right)^{1/2}}
+
\mathcal{O} \left( \left(1-e \right)^{1/2} \right),
\label{aavghighePR}
\end{equation}

\[
\widetilde{
\left\langle 
\left(
\frac{de}{dt} 
\right)^{\mathbb{Y}=0}
\right\rangle
}
= 
-
\frac
{5AL\mathcal{Q}}
{8 \sqrt{2} \pi m a^2 c^2 \left(1 - e\right)^{1/2}}
+
\mathcal{O} \left( \left(1-e \right)^{1/2} \right).
\]

\begin{equation}
\label{eavghighePR}
\end{equation}

%%%%%%%%%%%%%%%%% Figure
\begin{figure}
\centerline{
\psfig{figure=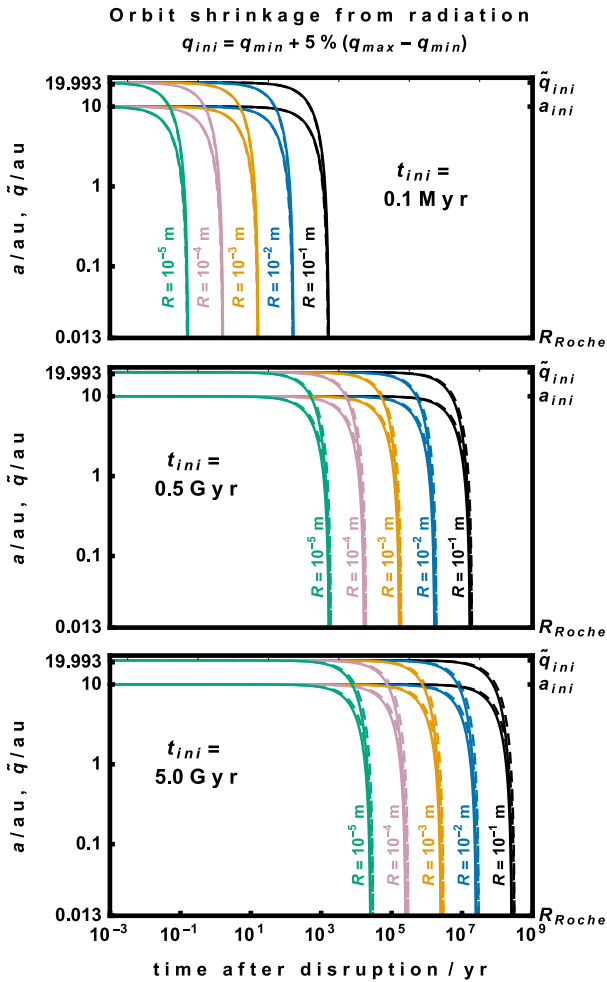,width=9.2cm} 
}
\caption{
Orbit shrinkage of various particle sizes (with radii $R$) from
WD radiation.  The three panels refer to different
WD ages ($t_{\rm ini}$) for a WD with a typical mass
of $0.6M_{\odot}$.  From top to bottom, these ages 
correspond to luminosities of about $22L_{\odot}$,
$2 \times 10^{-3} L_{\odot}$ and $1 \times 10^{-4} L_{\odot}$.
Plotted in each pair of similarly-coloured 
curves are the evolution of the semimajor axis ($a$; lower 
curve) and apocentre ($\tilde{q}$; upper curve).  {\it The eccentricity
of the orbit becomes zero only when those two lines converge}.
Solid lines are the result of numerical integrations of the full equation of motion 
(\ref{dadtavg}-\ref{dedtavg}) and 
the dashed lines are from our analytical approximations
(equations \ref{asol2}-\ref{asol3}). 
The initial semimajor axis was taken to be $a_{\rm ini} = 10$ au and
the pericentre to be $q_{\rm ini} = q_{\rm min} + 5\%(q_{\rm max} - q_{\rm min})$,
which corresponds to the orbital parameters of some rings 
formed in Paper I.  All orbits become compressed to entirely 
within the Roche radius.
}
\label{PRplots1}
\end{figure}
%%%%%%%%%%%%%%%%% Figure

\noindent{}Now suppose the debris ring is formed at the WD cooling age $t_{\rm ini}$, such that at this time each particle adopts semimajor axis and eccentricity values of $a_{\rm ini}$ and $e_{\rm ini}$.  We see from equation (\ref{qeq}) that

\begin{equation}
\lim_{e \rightarrow 1}
\left\langle 
\left(
\frac{dq}{dt} 
\right)^{\mathbb{Y}=0}
\right\rangle
=
0
.
\end{equation}

\noindent{}The semimajor axis and eccentricity change due to radiative effects, but on average the pericentre remains constant.  The eccentricity and semimajor axis evolution can thus be approximated as

\begin{equation}
a(t) \left[1 - e(t)\right] \approx a_{\rm ini}\left(1 - e_{\rm ini}\right)
.
\label{ehigh}
\end{equation}

\noindent{}Equation (\ref{ehigh}) may also be obtained 
by simultaneously solving the leading terms present in
equations (\ref{aavghighePR}) and (\ref{eavghighePR}).  
By using the relation in equation (\ref{ehigh}), 
we obtain

\begin{eqnarray}
\widetilde{
\left\langle 
\left(
\frac{da}{dt} 
\right)^{\mathbb{Y}=0}
\right\rangle
}
&\approx& 
-
\frac
{5AL\mathcal{Q} \sqrt{a}}
{8\sqrt{2} \pi m a_{\rm ini}^{3/2} c^2 \left(1 - e_{\rm ini}\right)^{3/2}}
.
\label{aavghighePR2}
\end{eqnarray}

In order to solve equation (\ref{aavghighePR2}) algebraically and explicitly for the semimajor axis, 
we express $L$ as in equation (\ref{lum}) but replace the power law exponent of $-1.18$ with $-6/5$, 
such that

\begin{equation}
L \approx L_0 \left(\frac{1}{10} + \frac{t}{\rm Myr}  \right)^{-6/5}
\end{equation}

\noindent{}where $L_0 = 3.26 L_{\odot} (M/0.6M_{\odot})$ and is hence constant.  Also, we express $A$ and
$m$ in terms of the body's radius ($R$) and density ($\rho$), as we are more interested in the
dependence on size than on mass.  The final result can be expressed
compactly through the following auxiliary variable

\[
\mathcal{K} \equiv 
\left(   
\frac
{75 \cdot 5^{1/5}}
{32 \cdot 2^{3/10} \pi}
\right)
\left(
\frac
{L_0 \mathcal{Q} {\rm Myr}}
{c^2 \rho R a_{\rm ini} \left(1 - e_{\rm ini}\right)^{3/2}}
\right)
\]

\begin{equation}
\times
\left[
\left(
1 + 
\frac{10t_{\rm ini}}{{\rm Myr}}
\right)^{-1/5}
-
\left(
1 +
\frac{10t}{{\rm Myr}}
\right)^{-1/5}
\right]
.
\end{equation}

\noindent{}The evolution of the semimajor axis is

\begin{equation}
a(t) = a_{\rm ini} - \mathcal{K} + \frac{\mathcal{K}^2}{4 a_{\rm ini}}
.
\label{asol2}
\end{equation}

\noindent{}Further, because the pericentre is constant, the apocentre reads

\begin{equation}
\tilde{q}(t) \approx a(t) - 2q_{\rm ini}.
\label{asol3}
\end{equation}

We plot the curves predicted from equation (\ref{asol2}) as 
dashed lines on Fig. \ref{PRplots1}.
The agreement with the true solution is excellent, for both 
the semimajor axis and apocentre, and for all cooling ages
relevant to WD pollution, at 
least to the precision we seek.  

Only physical solutions from equation (\ref{asol2})
are plotted.  After a body encounters the Roche radius, the time 
evolution is stopped.  Otherwise,
the semimajor axis would then unphysically increase after achieving a minimum.  
To stop the time evolution at the appropriate time, one can multiply equation 
(\ref{asol2}) by the appropriate Heaviside function.

Of particular interest is the {\it shrinking time}, $t_{\rm shr}$, which we define as the time taken for the particle's initial orbit to be compressed entirely within the WD Roche radius.  By using equation (\ref{asol2}), we find

\[
\left(1 + \frac{10\left(t_{\rm shr} + t_{\rm ini}\right)}{{\rm Myr}}\right)^{-1/5} 
= 
\left(1 + \frac{10t_{\rm ini}}{{\rm Myr}}\right)^{-1/5}
\]

\begin{equation}
-
\left(
\frac
{64 \cdot 2^{3/10} \pi}
{75 \cdot 5^{1/5}}
\right)
\left(
\frac
{c^2 \rho R q_{\rm ini}^{3/2}}
{L_0 \mathcal{Q} {\rm Myr}}
\right)
\left(
\sqrt{a_{\rm ini}}
\pm
\sqrt{2q_{\rm ini} + R_{\rm Roche}}
\right)
\label{shrtime}
\end{equation}

\noindent{}where $R_{\rm Roche}$ is the Roche radius of the
WD. From this equation $t_{\rm shr}$ may be solved for explicitly;
we adopted the (physical) solution with the lower sign.

\begin{figure}
\centerline{
\psfig{figure=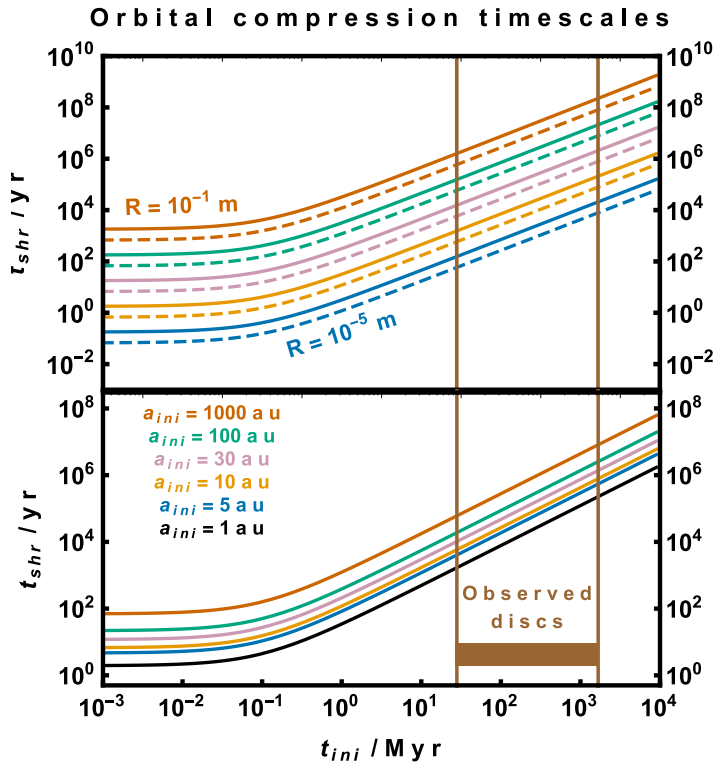,width=10.5cm} 
}
\caption{
The shrinking timescale $t_{\rm shr}$ as a function of WD cooling age $t_{\rm ini}$ for curves of different particle radii (upper panel) and initial semimajor axes (lower panel).  In all cases, a $\rho = 2$ g/cm$^3$ particle orbits a $0.6M_{\odot}$ WD.  In the upper panel, we adopt $a_{\rm ini} = 10$ au and $q_{\rm ini} = q_{\rm min} + 10\%(q_{\rm max} - q_{\rm min})$ (solid lines) and $q_{\rm ini} = q_{\rm min} + 5\%(q_{\rm max} - q_{\rm min})$ (dashed lines).  The similarly-coloured pairs of lines each corresponds to a different order of magnitude of particle radius.  In the lower panel, all particles have $R = 10^{-3}$ m and $q_{\rm ini} = q_{\rm min} + 5\%(q_{\rm max} - q_{\rm min})$.  Marked for reference as a thick brown bar and vertical brown lines is the range of cooling ages at which WD discs have been observed.  All curves in both panels demonstrate that the shrinking timescales are orders of magnitude shorter than the cooling times.
}
\label{PRplots2}
\end{figure}

We use equation (\ref{shrtime}) to generate Figure \ref{PRplots2},
which illustrates the dependencies of $t_{\rm shr}$ on $R$ and $a$.
For observational perspective we display the range of cooling
ages of known WD with debris discs ($\approx$ 30 Myr - 1.5 Gyr; 
\citealt*{faretal2009,giretal2012,beretal2014}) with a thick 
brown horizontal bar and vertical brown lines.
The steep increase in shrinking timescale during the first Gyr of
WD evolution is a reflection of the star's rapidly dwindling 
luminosity at early cooling ages.  Nevertheless, for all cooling
ages, the radiation is still strong enough to shrink and
circularize the orbits on timescales orders of magnitude smaller
than the WD cooling age.  Coincidentally, the largest particles
for which the Yarkovsky effect play no role ($\sim 10^{-1}$ m)
also approach the maximum particle size at which 
$t_{\rm shr} \ll t_{\rm ini}$, at least for $a_{\rm ini} = 10$ au.
Indeed, we do not yet have any observational constraints
on $a_{\rm ini}$; equations such as (\ref{shrtime}) may
help motivate likely values.

%%%%%%%%%%%%%%%%% Figure
%%%%%%%%%%%%%%%%% Figure

\section{Discussion}

With our results, we can now try to construct a consistent picture from Paper I about how tidal disruption of asteroids is followed by orbit compression.  A single asteroid is tidally disrupted into a collisionless ring with an unknown size distribution.  The orbit of the ring is equivalent to the orbit of the progenitor, whose eccentricity exceeds 0.98 and whose semimajor axis exceeds a few au.  A Main Belt-like asteroid would satisfy $a_{\rm ini} \approx 5$ au, a Kuiper Belt-like asteroid would satisfy $a_{\rm ini} \approx 30$ au, and an asteroid from a debris belt in a system like Fomalhaut would satisfy $a_{\rm ini} \approx 100$ au.  Because the ring is assumed to be collisionless, the bodies will not gravitationally perturb one another.  Every body will experience radiation forces in the form of Poynting-Robertson drag.

\subsection{Collisions upon contraction}

If the ring contains differently-sized bodies, then the effects of Poynting-Robertson-drag alone will cause the orbits of these bodies to contract at different rates.  As smaller bodies contract more quickly, their pericentres will precess more rapidly due to general relativity.  Consequently, collisions might occur.  Consider equation 19 of Paper I:

\begin{equation}
P_{\omega} \approx 0.15 \ {\rm Myr}
\left[
\frac{1 - e^2}
{1-0.999^2}
\right]
\left(
\frac{M_{\rm WD}}{0.6 M_{\odot}}
\right)^{-3/2}
\left(
\frac{a}{1 \ {\rm au}}
\right)^{5/2}
\label{GRinf}
\end{equation}

\noindent{}which illustrates that the general relativistic precession period $P_{\omega} \propto a^{5/2}$ and is also a function of $e$ and $M_{\rm WD}$.  Both these last two quantities remain nearly constant as the orbit contracts by a factor of tens to hundreds\footnote{Eventually, the eccentricity starts to decrease appreciably, potentially increasing the factor in square brackets by up to a factor of about 500.  Nevertheless, this eccentricity decrease does not occur until the particle has migrated nearly all the way to the Roche radius.}.  Therefore, suppose an asteroid on a $a = 10$ au orbit is broken up into large and small particles with radii $R_{\rm large}$ and $R_{\rm small}$.  The larger particle ring will continue to precess with a period of about 47 Myr while the smaller particle ring is shrunk; at $a = 1$ au, that small particle ring's precession period is just about 0.15 Myr.  Consequently, when $t_{\rm shr} \gtrsim P_{\omega}$, we can expect collisions to be significant; for small enough particles such that $t_{\rm shr} \ll P_{\omega}$, collisional effects would be negligible.

\subsection{Yarkovsky perturbations on chunky rings}

If there exist ring constituents that remain greater than about one metre in diameter over long-enough timescales, 
then the Yarkovsky effect will likely become the dominant contribution to the motion.  We mentioned this possibility
in Section 2, but now provide some quantification.  

Let $\mathbb{Q} \equiv Q_{\rm abs} \mathbb{I} + Q_{\rm ref} \mathbb{I} + k Q_{\rm yar} \mathbb{Y}$.
The components of $\mathbb{Q}$ are a function of spin state and orbital position, velocity and time. 
Although the simplified spherical internal heat model provided by \cite{broz2006} 
allows the time evolution of the semimajor axis or eccentricity to be expressed
(see Sections 2.3.2 and 2.3.3 of \citealt*{broz2006}, or \citealt*{vokrouhlicky1998}, 
\citealt*{vokfar1998} or \citealt*{vokrouhlicky1999}), that model itself might not be
applicable for the extremely high eccentricities considered in this paper.  For a particle on an 
eccentric orbit, the seasonal component of the Yarkovsky effect contains many forcing periods;
the amplitudes of those frequencies are functions of the orbital eccentricity.

Regardless of these difficulties, we now provide a zeroth-order demonstration of the potential
importance of the effect.  Each element of $\mathbb{Q}$ is strictly bounded between 0 and 2,
allowing us to bound the extent of the potential perturbation.  By assuming that the matrix elements
are independent of position and velocity, \cite{veretal2015a} obtained the resulting 
orbital element-based  unaveraged equations, as well as the averaged equations to leading order
in ($1/c$).  By expanding their equations A3 and A4 about $\epsilon=1-e$, $\epsilon \ll 1$, we obtain

\[
\widetilde{
\left\langle \frac{da}{dt} \right\rangle
}
=
\frac
{AL}
{4\pi c mna^2}
\left[   
\frac{1}{2\left(1-e\right)} + \frac{1}{4}
\right]
\]

\begin{equation}
\ \ \
\times
\left( \begin{array}{c} 
            \mathbb{Q}_{12} - \mathbb{Q}_{21}  \\
            \mathbb{Q}_{23} - \mathbb{Q}_{32}  \\
            \mathbb{Q}_{13} - \mathbb{Q}_{31}  \\
            \end{array}
            \right)
\cdot
\left( \begin{array}{c} 
            \cos{i}  \\
            \sin{i} \sin{\Omega}  \\
            \sin{i} \cos{\Omega}  \\
            \end{array}
            \right)
+\mathcal{O} 
\left( 1 - e \right)
\label{dadtavgYark},
\end{equation}

\[
\widetilde{
\left\langle \frac{de}{dt} \right\rangle
}
=
\frac
{AL}
{8\pi cmna^3}
\]

\begin{equation}
\ \ \
\times
\left( \begin{array}{c} 
            \mathbb{Q}_{12} - \mathbb{Q}_{21}  \\
            \mathbb{Q}_{23} - \mathbb{Q}_{32}  \\
            \mathbb{Q}_{13} - \mathbb{Q}_{31}  \\
            \end{array}
            \right)
\cdot
\left( \begin{array}{c} 
            \cos{i}  \\
            \sin{i} \sin{\Omega}  \\
            \sin{i} \cos{\Omega}  \\
            \end{array}
            \right) 
+\mathcal{O} 
\left( \sqrt{1-e} \right)
\label{dedtavgYark}.
\end{equation}

\noindent{}Comparing equations (\ref{dadtavgYark})-(\ref{dedtavgYark}) with 
(\ref{aavghighePR})-(\ref{eavghighePR}) reveals that 
(\ref{aavghighePR})-(\ref{eavghighePR}) are a
factor of about $1/c$ smaller than (\ref{dadtavgYark})-(\ref{dedtavgYark}), 
and can be neglected when
the Yarkovsky acceleration is ``turned on''.  Further, 
because the leading-order eccentricity
term of equation (\ref{dadtavgYark}) is of order $1/(1-e)$ whereas that of
equation (\ref{dedtavgYark}) is of order of just unity, the Yarkovsky 
acceleration might cause 
the orbit to shrink quickly at a nearly fixed eccentricity.

The off-diagonal terms will determine the sign of the expressions,
and hence whether the orbit shrinks or expands, circularizes or
becomes parabolic.  In fact, because the initial eccentricity is
so high, on average we might expect about half of the large chunks
in the ring to be ejected.  Given that the Yarkovsky acceleration
affects particles only above a certain size, the extra movement of these particles
with respect to the relative velocity of the constituents after
break-up might lead to collisions, which could produce a damping
effect.

Although we wish to obtain an expression similar to equation (\ref{asol2})
for the Yarkovsky effect, we cannot do so, but can make some progress
towards a solution.  In principle, we could follow 
the same procedure as in Section 5
to obtain an approximate closed formula for $a(t)$ due to the
leading-order terms in equations (\ref{dadtavgYark})-(\ref{dedtavgYark}).  
In the high eccentricity limit, we can relate $a$ and $e$ 
to one another and their
values at a particular cooling age because of the similar forms of equations 
(\ref{dadtavgYark})-(\ref{dedtavgYark}).   Dividing equation 
(\ref{dadtavgYark}) by equation (\ref{dedtavgYark}) and integrating yields

\begin{equation}
2a^2 = 2a_{\rm ini}^2 + 
\frac{1}{2} \left(e - e_{\rm ini}\right) + \ln{\left( \frac{1 - e_{\rm ini}}{1 - e} \right)}
.
\label{aeYark}
\end{equation}

\noindent{}However, we cannot go further and substitute equation (\ref{aeYark})
into equation (\ref{dadtavgYark}) and then integrate primarily because equations 
(\ref{dadtavgYark})-(\ref{dedtavgYark}) depend on the inclination and longitude
of ascending node (which are time-dependent) and the elements of
$\mathbb{Q}$ (which may be time-dependent).

\section{Conclusions}

We have addressed a problem in the formation of discs which orbit WDs at a distance of $1R_{\odot}$: how to shrink the $10-100$ au-scale extremely eccentric ($e > 0.98$) bound orbits of debris from tidally-disrupted asteroids. We demonstrated that this orbit compression readily occurs for particle sizes of $10^{-5} - 10^{-1}$ m on timescales many orders of magnitude shorter than the WD cooling age, due to WD radiation alone.  We provided explicit approximations for the shrinking time (equation \ref{shrtime}) as well as for the semimajor axis and apocentre evolution (equations \ref{asol2}-\ref{asol3}) during the orbit compression (the pericentre remains constant).  After the debris is perturbed close to or within the Roche radius, subsequent evolution is likely to be dictated by collisional and sublimative forces.

\section*{Acknowledgments}

We thank the referee, Miroslav Bro{\v z}, for reading our paper carefully and providing detailed comments, which have helped us improve and tighten the manuscript.  We also thank J.J. Hermes for useful discussions.
The research leading to these results has received funding from the European 
Research Council under the European Union's Seventh Framework Programme (FP/2007-2013) 
/ ERC Grant Agreements n. 320964 (WDTracer) and n. 282703 (NEOShield), as well as
from Paris Observatory's ESTERS (Environment Spatial de la Terre: Rechreche \& Surveilance) travel grants.

\label{lastpage}
\end{document}